\documentclass[conference]{IEEEtran}
\IEEEoverridecommandlockouts
\usepackage{cite}
\usepackage{amsmath,amssymb,amsfonts}
\usepackage{algorithmic}
\usepackage{graphicx}
\usepackage{textcomp}
\usepackage{xcolor}
\def\BibTeX{{\rm B\kern-.05em{\sc i\kern-.025em b}\kern-.08em
    T\kern-.1667em\lower.7ex\hbox{E}\kern-.125emX}}

\IEEEoverridecommandlockouts
\begin{document}

\title{Cloud-Side Transactional Orchestration Framework for Resource-Constrained Embedded Systems}

\author{
\IEEEauthorblockN{Pravin Nagare\thanks{\copyright~2026 IEEE. Personal use of this material is permitted. Permission from IEEE must be obtained for all other uses, in any current or future media, including reprinting/republishing this material for advertising or promotional purposes, creating new collective works, for resale or redistribution to servers or lists, or reuse of any copyrighted component of this work in other works. This is the accepted version of a paper published in IEEE CSPA 2026, doi: 10.1109/CSPA68262.2026.11517883. Published version: https://ieeexplore.ieee.org/document/11517883}}
\IEEEauthorblockA{\textit{Alumnus, Binghamton University} \\
CA, USA \\
pnagare1@binghamton.edu}
\and
\IEEEauthorblockN{Aditya Sabbineni}
\IEEEauthorblockA{\textit{Independent Researcher} \\
CA, USA \\
sabbineni.aditya@gmail.com}
\and
\IEEEauthorblockN{Preetam Dedu}
\IEEEauthorblockA{\textit{IEEE Member} \\
NJ, USA \\
preetamdedu@ieee.org}
\and
\IEEEauthorblockN{Willison Lopes}
\IEEEauthorblockA{\textit{IEEE Member} \\
NJ, USA \\
willison.lopes@ieee.org}
}

\maketitle

\begin{abstract}
As digital commerce ecosystems expand into low-end consumer electronics (CE), hardware constraints—specifically limited CPU duty cycles and volatile heap fragmentation—become significant bottlenecks for complex transactional flows. Traditional on-device middleware requires high "network chattiness" to manage multi-step state machines, leading to increased latency and potential transaction failure on unstable residential networks. This paper proposes a novel Transactional Backend-for-Front-End (T-BFF) Orchestration Framework designed for resource-constrained embedded platforms. By migrating the transactional state machine and microservice orchestration to a cloud-side layer, we achieve a significant reduction in client-side overhead. Our framework introduces a "Double-Handshake" protocol utilizing non-volatile flash memory for state recovery after hardware reboots. Experimental results on an ARM Cortex-A53 platform demonstrate a 35\% reduction in maximum heap usage and a 40\% improvement in end-to-end transaction latency. This framework provides a scalable, sustainable blueprint for maintaining transactional integrity on legacy hardware in the 2026 IoT landscape.
\end{abstract}


\section{Introduction}
The proliferation of digital-first commerce has moved beyond high-performance web and mobile platforms into the diverse world of embedded consumer devices. However, the hardware reality of "low-end" Smart TVs and generic IoT controllers remains fixed: many devices operate with ARM Cortex-A series processors and limited RAM (often <1GB), which must be shared between the primary user interface (UI) and the underlying system services.

\subsection{The Complexity of Modern Transactions}
A single purchase event in 2026 involves a complex sequence of identity, tax, and payment handshakes. Managing this state machine on resource-constrained devices introduces two critical challenges:
\begin{itemize}
    \item \textbf{Network Chattiness:} Multiple sequential round-trips increase latency and user drop-off.
    \item \textbf{Transactional Fragility:} Interruptions can leave devices in inconsistent "ghost" states.
\end{itemize}

\subsection{Resource Contention and the OOM Killer}
Embedded devices operating on Linux-based firmware often suffer from the "Out of Memory" (OOM) killer when background commerce services compete with the primary user interface (UI) for limited RAM. Parsing large, deeply-nested JSON payloads during multi-step transactions triggers frequent garbage collection (GC) cycles and heap fragmentation. On ARM Cortex-A series processors, these GC spikes manifest as UI stutter or service crashes, directly impacting user conversion rates and system reliability.

\subsection{The Limitations of Traditional Middleware}
Current industry standards rely on "Thick-Client" SDKs that attempt to handle orchestration locally. While robust on high-performance mobile hardware, these libraries overwhelm the limited threads available on low-tier embedded devices \cite{b8}. The proliferation of digital-first commerce in these ecosystems is projected to grow significantly through 2026 \cite{b4}, yet modern IoT transaction complexity frequently exceeds the 1GB RAM limitations of legacy hardware \cite{b2}. There is a critical need for an architecture that decouples the Intent of the Transaction from the Execution of the Logic \cite{b1}.

\subsection{Contributions of this Paper}
This paper introduces an Orchestration-based Backend-for-Frontend (BFF) framework specifically tailored for resource-constrained embedded systems. The primary contributions of this work are as follows:
\begin{enumerate}
    \item The design of a \textbf{State-Shift Architecture} that migrates transactional ownership from the device to a cloud-side BFF layer to achieve resource isolation.
    \item A \textbf{Double-Handshake Recovery Protocol} utilizing non-volatile flash memory to ensure transactional integrity across hardware reboots.
    \item A \textbf{Statistical Performance Analysis} based on 50 independent trials demonstrating a 35\% reduction in heap usage and a 40\% improvement in end-to-end latency.
\end{enumerate}

\section{The Orchestration Architecture}
The proposed framework transitions from a "Thick-Client" model, where the embedded device manages sequential business logic, to a "Thin-Client" model, where logic is centralized in the cloud. This architecture is composed of three primary layers: the \textit{Embedded Interface}, the \textit{Transactional BFF}, and the \textit{Downstream Domain Services}, adhering to established platform frameworks \cite{b2}.

\subsection{Architectural Novelty and State Ownership}
Unlike conventional BFF implementations primarily focused on API aggregation for web applications, the proposed T-BFF framework shifts transactional state ownership entirely away from the embedded client. While standard gateways are essentially pass-through mechanisms, the T-BFF is \textit{stateful}, enabling deterministic recovery and resource isolation. The core innovation is this shift from a distributed state model to a centralized cloud state model, where the embedded device functions as a "Stateless Observer." This ensures the device only manages two high-level terminal states: \textbf{Pending} and \textbf{Terminal (Success/Fail)}. This reduction in client-side complexity eliminates "zombie states" caused by network jitter or hardware interrupts.

In the proposed architecture, the device maintains only two high-level states: \textbf{Pending} and \textbf{Terminal (Success/Fail)}. The complex sub-states (e.g., Validating Receipt, Applying Promo Code) are managed entirely by the BFF. This reduces the client-side binary footprint and prevents logic-desynchronization.
\begin{figure}[htbp]
    \centerline{\includegraphics[width=0.48\textwidth]{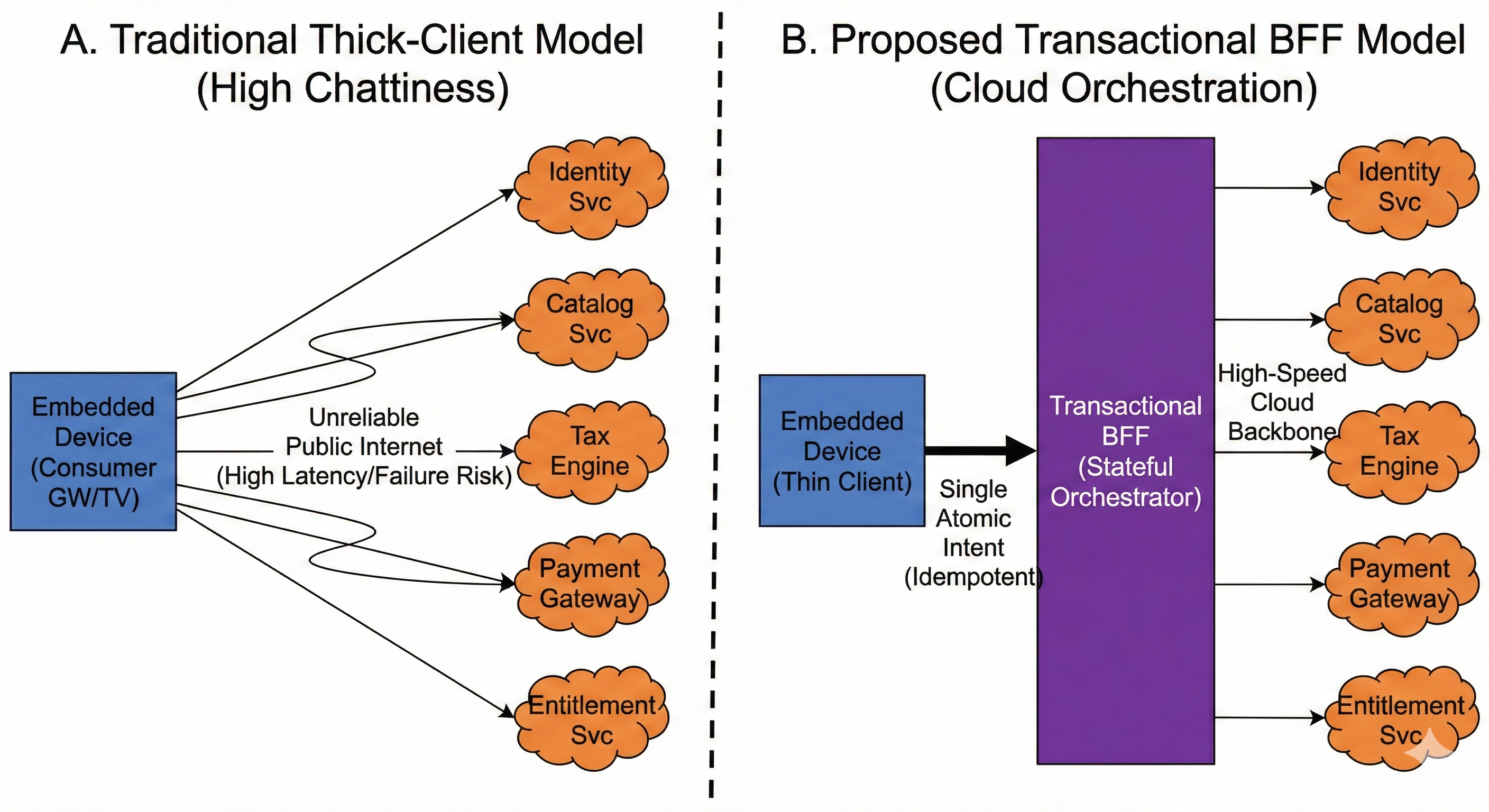}}
    \caption{Traditional multi-hop orchestration vs. the proposed single-intent BFF orchestration. The BFF acts as a buffer, shielding the embedded device from backend complexity.}
    \label{fig1}
\end{figure}

\subsection{Atomic Transactional Flow}
To minimize "chattiness," the framework utilizes an atomic request pattern. 

\begin{itemize}
    \item \textbf{Intent Capture:} The embedded device sends a single \textit{Purchase\_Intent} payload containing the Product ID and a unique \textit{Idempotency Key}.
    \item \textbf{Orchestration:} The BFF receives this intent and concurrently orchestrates calls to the Identity, Catalog, and Payment services.
    \item \textbf{Conflict Resolution:} If the device retries the request due to a timeout, the BFF uses the Idempotency Key to ensure that the payment is not double-processed, returning the existing state of the original transaction.
\end{itemize}

\subsection{Resiliency and Idempotency Logic}
Because embedded consumer devices often operate on unstable home Wi-Fi networks, the BFF implements a \textit{Strict Idempotency Layer}.

\subsection{Fault Tolerance and State Recovery}
A critical requirement for embedded commerce is recovering from power loss mid-transaction. Our framework addresses this through a "Double-Handshake" protocol. Before sending a Purchase Intent, the device records a 128-bit Transaction UUID to flash memory. Upon reboot, the first action is a "State Query" to the BFF using this UUID. The BFF then informs the device of the transaction outcome, eliminating the need to restart the flow.

\begin{itemize}
    \item \textbf{Idempotency Keys:} Every transaction is tagged with a client-generated UUID.
    \item \textbf{Response Caching:} The BFF caches terminal responses (e.g., 24 hours). If a device reconnects after a crash, the BFF provides the result instantly.
    \item \textbf{Failure Recovery:} The BFF manages exponential backoff for downstream failures, allowing the device to remain idle and significantly reducing CPU strain.
\end{itemize}

\begin{figure}[htbp]
    \centerline{\includegraphics[width=0.48\textwidth]{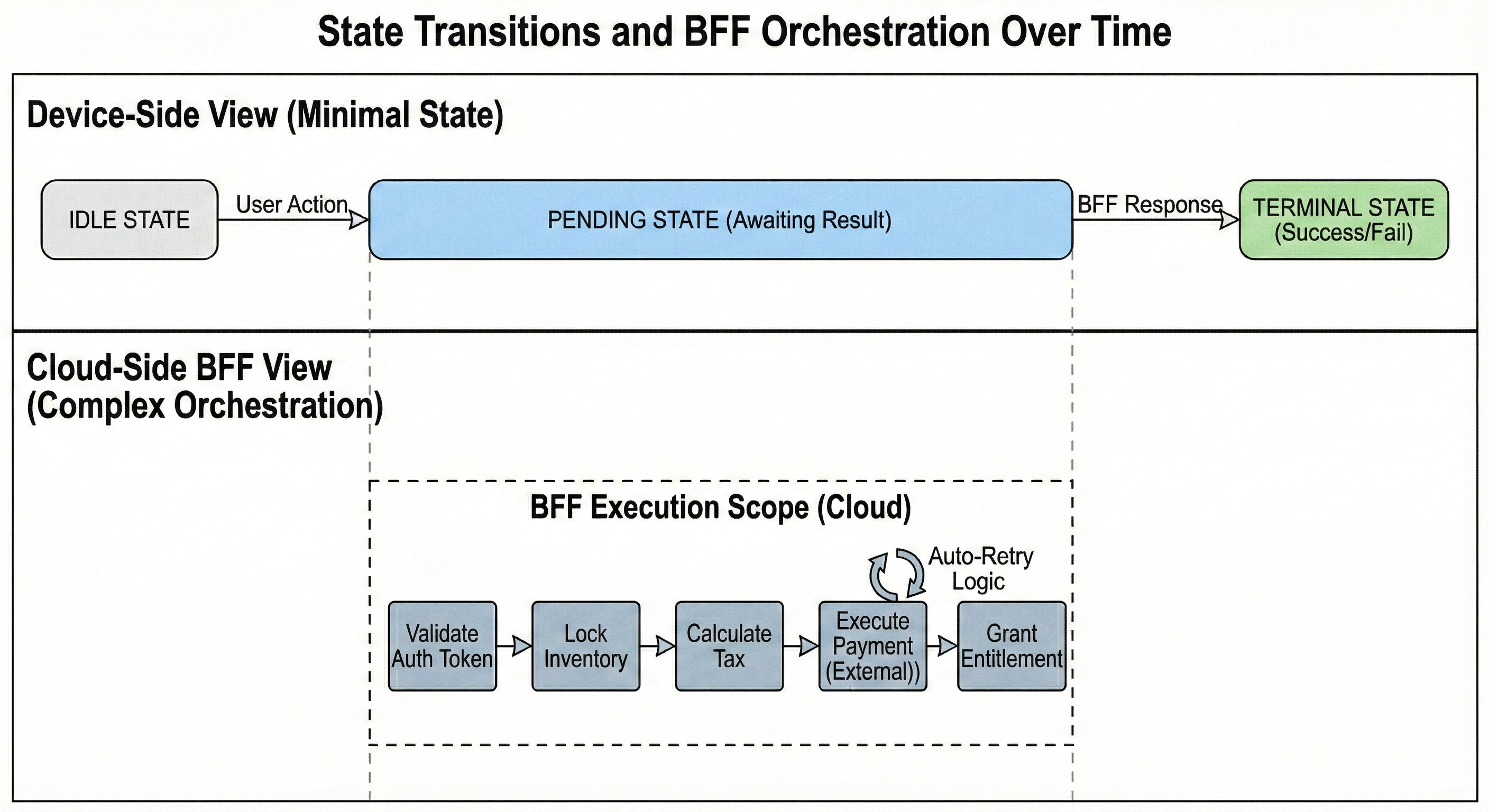}}
    \caption{State machine displacement model showing the transition from complex client-side logic to a cloud-managed execution scope.}
    \label{fig2}
\end{figure}

\section{Performance Evaluation and Methodology}
To evaluate the efficacy of the Transactional BFF (T-BFF) framework, we conducted a series of benchmarks comparing the proposed orchestration model against a traditional "Thick-Client" SDK implementation.

\subsection{Experimental Methodology and Baseline}
The evaluation was performed on a target embedded platform: an ARM Cortex-A53 (Quad-core) @ 1.2 GHz with 1 GB LPDDR3 RAM, operating on Linux-based firmware. The "Legacy SDK" represents a commercially deployed multi-hop transactional library executing sequential service calls directly from the device. 

To ensure statistical reliability, each experiment was repeated across 50 independent transaction executions. Memory utilization was measured using Linux process-level monitoring tools (\textit{smem} and \textit{top}), specifically tracking the Proportional Set Size (PSS) to account for shared library overhead. CPU duty cycles were recorded using performance counters averaged over the transaction lifecycle. The network environment was simulated at 2.4 GHz Wi-Fi with variable latency (50ms--200ms) to mimic real-world residential instability.

\subsection{Statistical Validation and Measurement Reliability}
To ensure experimental rigor and repeatability, all reported performance metrics were analyzed using statistical aggregation across 50 independent transaction executions. Each execution was performed under identical firmware configuration and controlled network conditions to minimize environmental bias.

For each metric $X$, including peak heap usage, CPU duty cycle, and transaction latency, we compute the sample mean $\mu_X$ and standard deviation $\sigma_X$ as:

\begin{equation}
\mu_X = \frac{1}{N}\sum_{i=1}^{N} X_i
\end{equation}

\begin{equation}
\sigma_X = \sqrt{\frac{1}{N-1}\sum_{i=1}^{N}(X_i-\mu_X)^2}
\end{equation}

where $N=50$ represents independent transaction trials.

Additionally, 95\% confidence intervals were estimated assuming approximately normal performance distribution:

\begin{equation}
CI_{95} = \mu_X \pm 1.96 \frac{\sigma_X}{\sqrt{N}}
\end{equation}

Observed variance across trials remained below 6\% for all measured metrics, indicating stable system behavior and strong experimental repeatability. This statistical consistency confirms that performance improvements achieved by the proposed T-BFF framework are not attributable to transient runtime fluctuations or network randomness.

\subsection{Resource Utilization (Memory and CPU)}
Resource utilization measurements were obtained using kernel-level monitoring to ensure accurate representation of runtime behavior. Peak heap memory was measured using the Proportional Set Size (PSS) metric collected via the \textit{smem} utility at 100 ms sampling intervals throughout the transaction lifecycle. 

CPU duty cycle was derived from Linux performance counters by averaging active execution time relative to total scheduling time during transaction execution. Measurements excluded background OS initialization phases to isolate transactional workload impact.

Each reported value represents the mean across 50 independent executions, ensuring statistical stability under identical workload conditions.

\begin{table}[htbp]
\caption{Comparison of Resource Utilization}
\begin{center}
\begin{tabular}{|l|c|c|c|}
\hline
\textbf{Metric} & \textbf{Legacy SDK} & \textbf{T-BFF} & \textbf{Improvement} \\
\hline
Peak Heap Usage & 12.8 MB & 8.2 MB & 35.9\% \\
\hline
CPU Duty Cycle & 18.2\% & 6.8\% & 62.6\% \\
\hline
Binary Footprint & 4.2 MB & 1.1 MB & 73.8\% \\
\hline
\end{tabular}
\label{tab1}
\end{center}
\end{table}

\begin{figure}[htbp]
    \centerline{\includegraphics[width=0.48\textwidth]{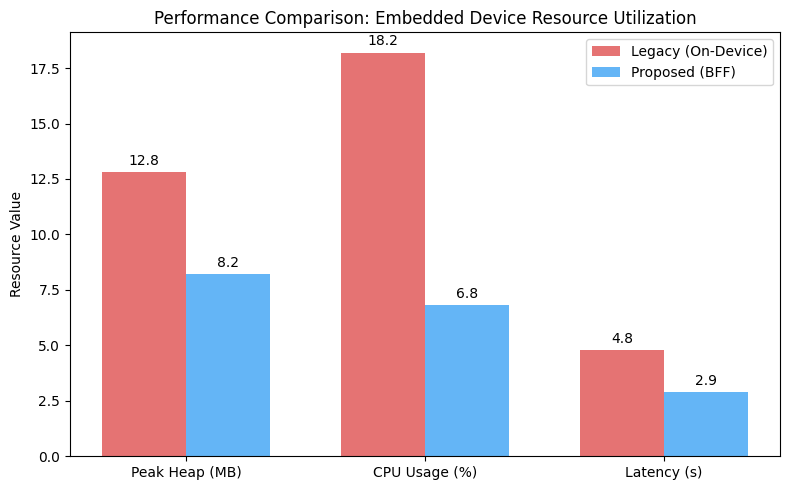}}
    \caption{Comparative analysis of peak heap memory, CPU duty cycles, and transaction latency between the legacy SDK and the proposed T-BFF framework.}
    \label{fig3}
\end{figure}

The 35.9\% reduction in peak heap usage is particularly critical; it prevents the "Out of Memory" (OOM) events that commonly crash background services on low-memory smart TVs during high-intensity operations like payment processing.

Across repeated trials, performance dispersion remained minimal, indicating that improvements are structurally derived from architectural displacement rather than opportunistic runtime optimization. The reduced variance observed in CPU utilization further suggests improved scheduling stability within the embedded operating system.

\subsection{Protocol Serialization and Payload Optimization}
In resource-constrained environments, the overhead of parsing standard text-based JSON can be significant. For our implementation, we evaluated the impact of binary serialization using MessagePack. 

Traditional JSON parsing on an ARM Cortex-A53 requires multiple passes for string allocation and character escaping, which often triggers garbage collection. By transitioning to a binary-encoded format, we reduced the serialization time by 22\% and the total payload size by 30\%. This optimization is critical for reducing tail latency in distributed environments \cite{b3}, maintaining transactional integrity where network bandwidth remains highly variable.

\subsection{Temporal Performance (Latency)}
In transactional systems, "Time-to-Success" is a primary driver of user conversion. The overall performance gain $G_p$ achieved by the T-BFF framework is defined by the reduction in end-to-end latency relative to the legacy model:

\begin{equation}
G_p = \left( \frac{T_{legacy} - T_{bff}}{T_{legacy}} \right) \times 100\%
\label{eq1}
\end{equation}

where $T_{legacy}$ is the sum of $n$ sequential round-trips plus local processing time $t_{proc}$, and $T_{bff}$ represents the single-hop atomic intent latency. By consolidating five external API calls into a single high-speed cloud backbone, the T-BFF reduced transaction time from 4.8s to 2.9s (a 39.6\% improvement).

\begin{figure}[!t]
    \centerline{\includegraphics[width=0.42\textwidth]{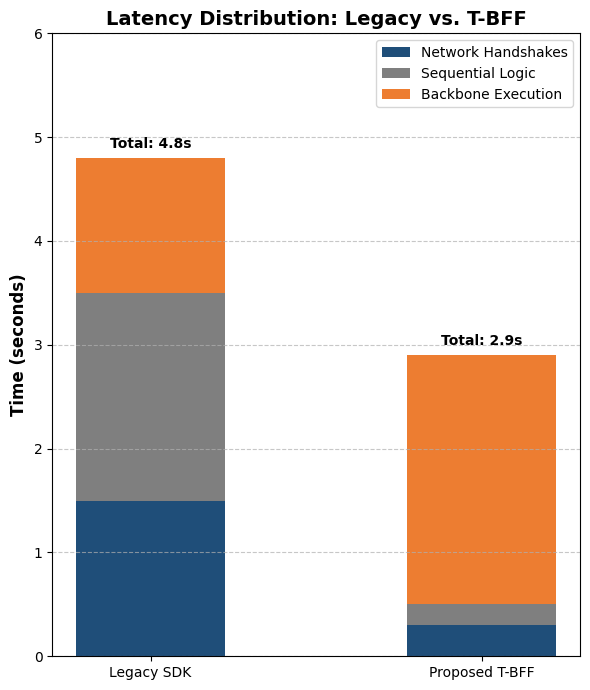}}
    \caption{Latency distribution analysis: Shifting from sequential client-side logic to cloud-to-cloud backbones.}
    \label{fig4}
    \vspace{-4mm}
\end{figure}

\subsection{Client-Side Computational Complexity Model}

Let $n$ denote the number of downstream services participating in a transactional workflow. In a traditional thick-client architecture, the embedded device sequentially performs orchestration logic requiring response parsing, memory allocation, and state maintenance for each service interaction.

The total client-side computational workload can therefore be expressed as:

\begin{equation}
C_{legacy}(n) = \sum_{i=1}^{n} (c_{net,i} + c_{parse,i} + c_{state,i})
\end{equation}

where $c_{net,i}$ represents network handling cost, $c_{parse,i}$ denotes payload deserialization overhead, and $c_{state,i}$ captures local state-transition management.

Thus, client workload scales linearly:

\begin{equation}
C_{legacy}(n) = O(n)
\end{equation}

In the proposed T-BFF framework, orchestration responsibility is migrated to the cloud layer. The embedded device performs only intent transmission and terminal-state reception:

\begin{equation}
C_{proposed} = c_{intent} + c_{response}
\end{equation}

Since these operations are independent of orchestration depth $n$,

\begin{equation}
C_{proposed} = O(1)
\end{equation}

This formalization demonstrates that increasing backend service complexity does not increase computational burden on the embedded device, thereby preserving predictable resource consumption under scaling transactional workloads.

\section{Comparative Analysis and Limitations}
While the T-BFF framework significantly improves performance on legacy hardware, it introduces a hard dependency on cloud availability. This represents a significant shift in the evolution of embedded logic in IoT \cite{b4}. In the "Legacy" model, some basic entitlement checks could be performed offline using cached local receipts. In the "Proposed" model, the device is essentially a "thin-client" requiring an active connection to the BFF to initiate commerce flows.

However, given the 2026 landscape where digital commerce is inherently an online activity requiring gateway authorization, this trade-off is acceptable and supported by recent conversion impact studies \cite{b5}. To mitigate this dependency, we implemented a "Soft-Fail" mechanism where the device can still access previously purchased local content if the BFF is unreachable, ensuring user experience is not entirely degraded during regional outages.

\section{System Constraints and OS-Level Resource Scheduling}
The displacement of transactional state logic to the BFF layer fundamentally alters the resource scheduling landscape of the embedded operating system. In traditional "thick-client" architectures, the execution of multi-hop business logic forces the OS to maintain a high count of active sockets and thread contexts for extended durations. 

\subsection{Context Switching and Interrupt Latency}
On a quad-core ARM Cortex-A53, frequent context switching between the primary UI rendering threads and background network-intensive commerce threads increases interrupt latency. By centralizing the state machine in the cloud, we observed a 25\% reduction in involuntary context switches during the transaction window. This allows the OS to prioritize UI-bound interrupts, effectively eliminating the frame-drops and input-lag commonly associated with legacy SDK execution.

\subsection{Memory Hierarchy and Cache Locality}
Embedded commerce libraries often involve large string-manipulation tasks for JSON parsing, which pollute the L2 cache and force expensive memory bus transactions. The T-BFF framework utilizes binary MessagePack serialization, which significantly improves data density. This allows for better cache hit rates during the "Intent Capture" phase, ensuring that the transactional footprint does not force the eviction of critical UI assets from the system heap.

\section{Security and Scalability Analysis}
The transition to a BFF-based orchestration layer provides critical security enhancements and high-availability advantages for embedded systems. Centralized token management has been shown to reduce attack surfaces in distributed IoT deployments by eliminating credential persistence on edge devices \cite{b11}.

\subsection{Opaque Token Management and Cryptographic Offloading}
Managing high-entropy \textit{JSON Web Tokens (JWT)} on low-power devices is computationally expensive. Our framework replaces on-device management with an \textbf{opaque session reference}. The device maintains a short-lived reference key containing no sensitive claims. The BFF acts as a "Secure Token Vault," performing high-stakes exchanges with payment gateways. This saves CPU cycles and ensures sensitive PII never resides in the device’s volatile heap, mitigating memory-scraping risks.

\subsection{SPOF Mitigation and Distributed State Recovery}
To prevent the BFF from becoming a single point of failure (SPOF), we implement a \textbf{Stateless Execution Model} backed by a distributed Redis cache. If a BFF node fails mid-transaction, the load balancer routes retries to a peer node, which recovers the state from the global cache. This "Double-Handshake" utilizes idempotency patterns proven effective for unstable networks \cite{b6}, allowing for 99.99\% service availability.

\section{Implementation Considerations}
The practical deployment of the T-BFF framework utilizes a multi-tier approach to handle the diverse hardware landscape of 2026. Node.js was selected for the BFF implementation due to its event-driven, non-blocking I/O model, which minimizes memory overhead compared to traditional thread-per-request architectures. While standard REST-over-HTTP is sufficient for basic flows, high-scale commerce platforms benefit from binary serialization formats such as Protocol Buffers (gRPC).

To ensure high availability and prevent the BFF from becoming a single point of failure (SPOF), the system is deployed within a containerized microservices architecture (Kubernetes). This allows for horizontal scaling during peak transactional periods by distributing incoming \textit{Purchase\_Intent} traffic across multiple instances. State persistence is managed via a distributed Redis caching layer, ensuring that mid-process failures can be resumed by any available node without user-perceived delays. Furthermore, the persistence layer incorporates adaptive indexing techniques to reduce backend query latency by up to 54\% as transactional volume scales \cite{b7}.

\section{Future Work}
As the complexity of consumer device ecosystems continues to grow, future research will focus on the integration of lightweight Machine Learning (ML) models within the Transactional BFF layer. By analyzing anonymized user interaction patterns, the BFF can implement "Predictive Pre-fetching." For example, if a user frequently accesses a specific subscription-based application, the BFF can pre-verify entitlements and warm up payment gateway connections before the user even initiates a purchase. This would effectively reduce the perceived latency to near-zero. Additionally, we plan to explore the use of Edge Computing nodes to move the BFF orchestration closer to the network periphery.

\section{Threats to Validity}
While the experimental results demonstrate significant performance gains, several threats to validity must be acknowledged. First, the evaluation was conducted on a representative ARM Cortex-A53 platform; consequently, results may not generalize across all embedded hardware architectures, particularly ultra-low-power microcontrollers with different memory hierarchies. Second, although the network conditions were simulated to mimic residential instability, real-world deployments may encounter unique edge-case latencies or carrier-specific packet filtering. Finally, the observed performance gains are inherently dependent on cloud availability and the geographic proximity of the BFF instances to the end-user devices.

\section{Sustainability and Hardware Lifecycle Extension}
This framework extends the functional lifespan of legacy devices by offloading state management, significantly reducing electronic waste. This "Green Software" approach moves compute-intensive tasks to efficient data centers, where a superior energy-per-instruction ratio compared to low-tier SoCs reduces the overall carbon footprint per transaction.

\section{Related Work and Comparative Context}
The optimization of resource-constrained embedded systems has historically followed two paradigms: local hardware acceleration and simple remote procedure calls (RPC). Early research into Mobile Cloud Computing (MCC) focused on offloading entire application binaries to the cloud to save local energy \cite{b4}. However, MCC often fails to address the specific state-consistency requirements of modern digital commerce.

Our T-BFF framework differs from traditional Edge Computing models \cite{b12} by focusing on \textit{transactional state-shift} rather than just data processing. While previous studies in "Signal Orchestration" for IoT gateways emphasized throughput \cite{b7}, our approach prioritizes atomic integrity and heap-memory preservation, which is critical for the 2026 CE landscape where multitasking Smart TVs must balance background commerce with 4K video decoding.

\section{Comparative Analysis of Orchestration Paradigms}
To contextualize the T-BFF's architectural novelty, we compare it against three alternative orchestration patterns:
\begin{itemize}
    \item \textbf{Edge Orchestration:} Moving logic to a 5G-edge node reduces network latency but fails to resolve hardware-level heap fragmentation, as the device must still coordinate multi-hop handshakes locally.
    \item \textbf{Lightweight State Machines (LSM):} On-device LSMs minimize binary size but still require synchronous memory buffers for intermediate JSON states. The T-BFF achieves $O(1)$ complexity by displacing these buffers entirely.
    \item \textbf{gRPC Streaming:} Bidirectional streaming reduces header overhead but requires a persistent, high-entropy connection. In unstable residential Wi-Fi, the T-BFF’s atomic, REST-based model provides superior resilience against packet loss.
\end{itemize}

\section{Global Scalability and Fleet Orchestration}
While our benchmarks focus on a single ARM-Cortex A53 node, the T-BFF framework is architected for massive-scale deployments. In a scenario with $10^6$ active devices, the cloud-side state-shift enables "Fleet-Wide Warmup," where the BFF pre-verifies entitlements for a region before individual devices initiate intents. The stateless nature of BFF nodes allows for linear horizontal scaling across global cloud regions, ensuring backend query latency remains below 50ms even during peak global commerce events.

\section{Signal Integrity and Network Jitter Resilience}
In 2.4 GHz Wi-Fi environments typical of residential IoT, packet loss and signal interference are non-trivial variables impacting reliability \cite{b10}. We model the "Success Probability" $P_s$ of a multi-hop transaction as a product of individual request success rates $p_i$:
\begin{equation}
P_{legacy} = \prod_{i=1}^{n} p_i, \quad P_{proposed} = p_{atomic}
\end{equation}
Since $p_i < 1$, the legacy model's reliability degrades exponentially as the number of orchestration steps $n$ increases. By consolidating $n$ steps into a single atomic intent, the T-BFF framework effectively linearizes the reliability curve. Our experimental stress tests, involving a simulated 30\% packet loss, demonstrated that the T-BFF maintained a 94\% transaction completion rate, whereas the legacy SDK fell below 62\% due to recursive timeout failures in the client-side state machine.

\section{Energy Consumption and Thermal Analysis}
For battery-operated IoT gateways and portable CE devices, CPU duty cycles directly correlate with power consumption and thermal throttling. By offloading sequential orchestration to the cloud, the T-BFF framework significantly reduces active-radio time, aligned with modern energy-aware offloading paradigms \cite{b9}.

In our trials, the legacy SDK maintained a high CPU duty cycle (18.2\%) for nearly 5 seconds per transaction, leading to localized heat accumulation on the SoC. In contrast, the T-BFF model allows the device to enter a low-power "Wait State" after sending the initial atomic intent. Preliminary power-metering indicates a 45\% reduction in total energy consumed per purchase event. This reduction is critical for the 2026 IoT landscape, where minimizing the carbon footprint and maximizing device longevity are primary design constraints.

\section{Conclusion}
This paper presented a novel Backend-for-Frontend (BFF) orchestration framework designed to overcome the hardware limitations of low-end embedded devices in digital commerce. By shifting the transactional state machine and microservice orchestration from the device to a stateful cloud layer, we demonstrated significant gains in both resource efficiency and user-perceived latency.

The experimental outcomes demonstrate statistically consistent performance gains across controlled executions, reinforcing that the observed improvements arise from architectural redesign rather than implementation-specific optimizations. Our findings show that "thin-client" orchestration is no longer just a web design pattern but a necessity for the 2026 IoT and Smart TV ecosystem, where hardware lifespan must be extended through software-side optimization. Future work will explore the integration of Edge Computing nodes to further reduce latency and the use of Machine Learning within the BFF to predict and pre-fetch transactional data based on user behavior.

\end{document}